\documentclass[twocolumn]{revtex4}

\usepackage{graphicx}
\usepackage{xcolor}
\usepackage[normalem]{ulem}
\usepackage{multirow}
\usepackage{amsmath}
\newcommand*{\blue}{\color{blue}}

\begin{document}

\title{Is the $H_0$ tension suggesting a 4th neutrino generation?}

\author{S. Carneiro$^{1,2}$, P. C. de Holanda$^{1}$, C. Pigozzo$^{2}$ and F. Sobreira$^{1}$}

\affiliation{$^1$Instituto de F\'{\i}sica Gleb Wataghin - UNICAMP, 13083-970, Campinas, SP, Brazil\\$^2$Instituto de F\'{\i}sica, Universidade Federal da Bahia, 40210-340, Salvador, Bahia, Brazil}

\date{\today}

\begin{abstract} 
Flavour oscillations experiments are suggesting the existence of a sterile, $4$th neutrino generation with a mass of an eV order. This would mean an additional relativistic degree of freedom in the cosmic inventory, in contradiction with recent results from the Planck satellite, that have confirmed the standard value $N_{\mathrm{eff}} \approx 3$ for the effective number of relativistic species. On the other hand, the Planck best-fit for the Hubble-Lema\^itre parameter is in tension with the local value determined with the Hubble Space Telescope, and adjusting $N_{\mathrm{eff}}$ is a possible way to overcome such a tension. In this paper we perform a joint analysis of three complementary cosmological distance rulers, namely the CMB acoustic scale measured by Planck, the BAO scale model-independently determined by Verde {\it et al.}, and luminosity distances measured with JLA and Pantheon SNe Ia surveys. Two Gaussian priors were imposed to the analysis, the local expansion rate measured by Riess {\it et al.} and the baryon density parameter fixed from primordial nucleosynthesis by Cooke {\it et al.}. For the sake of generality and robustness, two different models are used in the tests, the standard $\Lambda$CDM model and a generalised Chaplygin gas. The best-fit gives $N_{\mathrm{eff}} \approx 4$ in both models, with a Chaplygin gas parameter slightly negative, $\alpha \approx -0.04$. The standard value $N_{\mathrm{eff}} \approx 3$ is ruled out with $\approx 3\sigma$.
\end{abstract}

\maketitle

\section{Introduction}

The panorama on neutrino flavour oscillation experiments is very robust. Data from different experimental setups converge into a concise explanation, in which neutrino flavour oscillations are driven by two large and one small mixing angles and two hierarchical mass differences~\cite{Esteban:2016qun}. Such framework provides a precise prediction on flavour transitions of atmospheric, solar, reactor and accelerator neutrinos, in an energy range that varies from sub-MeV to several GeV, and distances that vary from few meters to astrophysical distances. These predictions have been corroborated by different experimental results on the last decades.

However, experiments that find neutrino flavour conversion signals that are not easily accommodated in the 3-neutrino mixing framework are piling up. More than 15 years ago the LSND experiment~\cite{Aguilar:2001ty} observed an appearance of electron anti-neutrinos in a muon anti-neutrinos flux, which if explained through mass-driven flavour oscillations would suggest a mass scale incompatible with others oscillation experiments results. Recently MiniBoone~\cite{Aguilar-Arevalo:2018gpe} confirmed the main features of LSND results, both in neutrino and anti-neutrino channels, strengthening the hypothesis that there is a fourth neutrino family, which does not couple with weak gauge bosons (hence, sterile neutrinos), but participates in flavour neutrino oscillations with a mass scale of order $\sim 1$ eV. 

Although the above mentioned results can be well explained by a fourth neutrino family, it seems that they are incompatible with disappearance experiments, such as Minos/Minos+~\cite{Adamson:2017uda}, NEOS~\cite{Ko:2016owz} and Daya Bay~\cite{An:2016luf} (see for instance~\cite{Dentler:2018sju} for a comprehensive comparison between experiments). Therefore, assuming that these experimental results should be explained by new physics on neutrino sector, it seems that such new physics would have to go beyond the simple addition of an extra neutrino family. As stated in~\cite{Liao:2018mbg}, the neutrino sector seems quite baroque.

Nevertheless, most of the solutions proposed to accommodate all oscillation neutrino experiments results would add an extra degree of freedom in the relativistic species that would be produced in the early universe. It is then worthwhile to revisit the cosmological results on this subject \cite{Hamann,Giusarma,bernal,suecos}. In the present contribution we analyse two distance rulers that are sensitive to the number of relativistic species, namely the CMB and BAO acoustic scales, complemented by SNe Ia luminosity distances observations and by the current priors on the local expansion rate and baryonic density.

The paper is organised as follows. In the next section we discuss why the tension between Planck and HST measurements of $H_0$ can be alleviated with a higher $N_{\mathrm{eff}}$ value. In section 3 we describe the tests to be performed, and in sections 4 and 5 we show the results of our joint analysis. In section 6 some conclusions are outlined.

\section{The acoustic horizon}

The acoustic horizon, given by
\begin{equation}
r_s(z) = \int_{z}^{\infty} \frac{c_s}{H(z')} dz',
\end{equation}
has two important values in the context of cosmological data. When we are dealing with the CMB acoustic scale $\theta_*$, the acoustic horizon is evaluated at the redshift of last scattering ($z_{*} \approx 1090$), so that $r_*\equiv r_s(z_*)$. In the case of BAO, the acoustic horizon is evaluated at the drag epoch  ($z_{d} \approx 1060$),  which we will refer to as $r_d\equiv r_s(z_d)$. The sound speed is given by
\begin{equation}
\frac{c_s}{c} = \left[ 3 + \frac{9\Omega_{b0}}{4\Omega_{\gamma0}} (1+z)^{-1} \right]^{-1/2},
\end{equation}
and the Hubble-Lema\^itre function of the spatially-flat standard model by
\begin{equation}
H(z) = H_0 \sqrt{(1-\Omega_{m0}) + \Omega_{m0} (1+z)^3 + \Omega_{R0} (1+ z)^4}.
\end{equation}
In the above expressions, $\Omega_{m0}=\Omega_{dm0} + \Omega_{b0}$ and $\Omega_{R0} = \Omega_{\nu0} + \Omega_{\gamma0}$ are, respectively, the density parameters of total matter (dark matter + baryons) and radiation (neutrinos + photons), and $H_0 = 100\, h$ km/s Mpc$^{-1}$ is the Hubble-Lema\^itre parameter. The radiation density parameter can be expressed as
\begin{equation}
\Omega_{R0} = \Omega_{\gamma0} \left[ 1+ 0.68\, (N/3)  \right],
\end{equation}
where $N$ is the number of neutrinos species. In a rough estimation, neglecting the contribution of the baryonic and dark sectors for $z \gg 1$, and taking the observed $\Omega_{\gamma0} h^2 = 2.47 \times 10^{-5}$ \cite{Planck}, we have
\begin{equation}
r_d^h \propto \frac{h}{\sqrt{2.47 \left[ 1 + 0.68\, (N/3) \right]}},
\end{equation}
where $r_d^h \equiv r_d h$. Let us consider a hypothetical observational probe of the acoustic scale, and let $\tilde{h}$ be the value obtained when the number of species is fixed in $\tilde{N} = 3$. For an arbitrary $N$, the same probe will give a Hubble-Lema\^itre parameter $h$ such that
\begin{equation}
\frac{N}{3} = 2.47 \left( \frac{h^2}{\tilde{h}^2} \right) - 1.47.
\end{equation}
Using for $\tilde{h}$ the Planck value $\tilde{h} = 0.68$ \cite{Planck}, and for $h$ the local value $h = 0.73$ \cite{Riess}, it follows that $N \approx 4.1$.

\section{Standard rulers}


We will consider two standard rulers in our analysis. The first is given by the position of the first peak in the CMB spectrum of anisotropies, more precisely the angular scale
\begin{equation}
{\theta_*} =  \frac{r_*}{D_A(z_{*})},
\end{equation}
where $D_A$ is the comoving angular diameter distance to the last scattering surface,
\begin{equation}
D_A(z_*) = \int_0^{z_{*}} \frac{c}{H(z)}dz.
\end{equation}
Its observed value is  $100\,\theta_{*} =  1.04109 \pm {0.00030}$ \cite{Collaboration:2018va}. The second ruler comes from BAO observations, and can be encompassed, in an approximately model-independent way, in the acoustic horizon derived by Verde {\it et al.}, $r_d^h = 101.2 \pm 2.3$ \cite{Verde}.
We will complement the analysis by fitting the luminosity distances to supernovae Ia of the JLA compilation \cite{sne:betoule14}. Compared to other surveys, it has the advantage of allowing the light-curve recalibration with the model under test. Although it was also used to derive the Verde {\it et al.} acoustic horizon at the drag epoch \cite{Verde}, this fitting is insensitive to $N_{\mathrm{eff}}$, and will be used for better constraining the matter density. Anyway, in order to control the effect of a double counting, we will also use the Pantheon SNe Ia compilation \cite{Scolnic:2017caz} in the analysis, which contains supernovae not used in the Verde {\it et al.} fitting. As Gaussian priors of our analysis, we will take the Riess {\it et al.} local value of the Hubble-Lema\^itre parameter  \cite{Riess}, $h = 0.7348\pm0.0166$, and the Cooke {\it et al.} value for the baryonic density parameter, $\Omega_{b0} h^2 = 0.02226 \pm 0.00023$, which comes from nucleosynthesis constraints \cite{Cooke}.

For the sake of generality, our tests will be performed with two different models. The first is the standard model, for which the indication of a 4th neutrino generation will already be manifest. The robustness of this possibility will be verified by testing an extension of the standard model given by the generalised Chaplygin gas \cite{gCg1,gCg2,gCg3,gCg4,gCg5,gCg6,gCg7,gCg8}, with a Hubble function given, with the addition of radiation, by
\begin{eqnarray}
\left[ \frac{H(z)}{H_0} \right]^2 &=&  \left[ (1 - \Omega_{m0}) + 
      \Omega_{m0} (1 + z)^{3(1 + \alpha)} \right]^{1/(1 + \alpha)} \nonumber \\ &+& \Omega_{R0}\, (1 + z)^4.
\end{eqnarray}
In the binomial expansion of the brackets, we have a leading term $\Omega_{m0} (1+z)^3$, which shows that, for the present purpose of background tests, the baryonic content can be absorbed in the above defined gas. For $\alpha = 0$ we recover the standard $\Lambda$CDM model. Perturbative tests are outside the scope of this paper, but let us comment that, although the adiabatic generalised Chaplygin gas is ruled out by the observed matter power spectrum \cite{waga}, non-adiabatic versions with negative $\alpha$ present a good concordance when tested against background and LSS observations \cite{wands21,wands22,wands23,non-adiabatic1,non-adiabatic2,wang,cassio}.

\section{Joint analysis and results}

On the basis of Bayesian Statistics, we defined the joint log-likelihood as a function of the parameter array {\bf p}, adding to the CMB log-likelihood,
\begin{equation}
\log\mathcal{L}_{\mathrm{CMB}}({\bf p}) = -0.5 \left(\frac{100\theta_*({\bf p})-1.04109}{ 0.00030}\right)^2,
\end{equation}
a log-likelihood for $r_d^h$,
\begin{equation}
\log\mathcal{L}_{\mathrm{BAO}}({\bf p}) = -0.5 \left(\frac{r_d^h({\bf p})-101.2}{2.3}\right)^2,
\end{equation}
and the log-likelihood of supernovae,  
\begin{eqnarray} \label{chi2_sne}
&\log\mathcal{L}_{\mathrm{SNe}}({\bf p}) =  &\nonumber  \\
&-0.5 \sum (\mathrm{\mathbf{m_{B}}}-\mathrm{\mathbf{m_{B}^{mod}}})^{\mathrm{\mathbf{T}}} (\mathrm{\mathbf{C^{-1}_{SN}}})(\mathrm{\mathbf{m_{B}}}-\mathrm{\mathbf{m_{B}^{mod}}}). &
\end{eqnarray}. 

For the Chaplygin gas the set of free cosmological parameters were ${\bf p_c} = \{H_0,\Omega_{b0}h^2,\Omega_{dm0}h^2,\alpha, N_{\rm{eff}}\}$, with free nuisance parameters due to corrections on SNe light-curves, ${\bf p_s} = \{\alpha_s,\beta_s,\mathcal{M}_B,\Delta_M\}$ for JLA SNe likelihood \cite{sne:betoule14,guy10} or ${\bf p_s} = \{\mathcal{M}_B\}$ for Pantheon SNe likelihood \cite{Scolnic:2017caz}, so that ${\bf p} = \{{\bf p_c},{\bf p_s}\}$.  

The supernovae theoretical apparent magnitude $m_{B}^{\mathrm{mod}}$ is written as 
\begin{equation}\label{mmod}
m_{B}^{\mathrm{mod}} = 5 \log_{10} d_L(z_{\mathrm{CMB}}, z_{\mathrm{hel}}) - \alpha_s X_1 + \beta_s \mathcal{C} + \mathcal{M}_B ,
\end{equation}
and so JLA supernovae light-curves are standardised along with cosmological parameters of the tested model, finding the best stretch ($\alpha_s$) and color ($\beta_s$) corrections, and also fitting the absolute magnitude $\mathcal{M}_B$ with a step $\Delta_M$ for more massive host galaxies. Even the covariance matrix $\mathbf{C_{SN}}$ is a function of $\alpha_s$ and $\beta_s$. For the Pantheon sample, $X_1$ and $\mathcal{C}$ values are not available, and the apparent magnitude and covariance matrix already calibrated for the standard $\Lambda$CDM model is given, so that it is allowed to adjust only the absolute magnitude $\mathcal{M}_B$.

We explored the parameter space via the PyMultiNest \cite{multinest01,multinest02,multinest03,pymultinest} module for Python,  setting 1500 live points and {\it `parameter'} sampling efficiency. Besides the Gaussian priors previously mentioned, all other parameters had uniform priors presented on Table \ref{tab:priors}.
\begin{table}[th!]
	\centering
	\begin{tabular}{c c }
	\hline
		Parameter			&	Uniform Prior			\\ \hline 
    $N_{\mathrm{eff}}$ & $\mathcal{U}$[0.05,10.00]\\
	$\Omega_{dm0}h^2$	& $\mathcal{U}[0.001,1.000]$		\\ 
	$\alpha$	   	&  $\mathcal{U}[-0.99,2.00]$			\\
	\hline 
	$\alpha_s$	   	&  $\mathcal{U}[0,1]$			\\
	$\beta_s$	   	&  $\mathcal{U}[0,4]$			\\
	$\mathcal{M}_B$	   	&  $\mathcal{U}[-22,-16]$		\\ [1ex]
	$\Delta_M$	   	&  $\mathcal{U}[-1,1]$		\\ 
	\hline
	\end{tabular}
	\caption{Lower and higher limits of the flat priors used in the analysis. The  last four rows are related to supernovae nuisance parameters.}
	\label{tab:priors}
	\end{table}
The results of our joint analysis are summarised in Fig. 1. 
Previous results \cite{cassio} with $N_{\mathrm{eff}}=3.046$, and JLA dataset only, favoured negative values of $\alpha$, which is not obtained in the present scenario. Also, even in the standard model case{\blue ,} a 4th neutrino generation is suggested by the data. The $2\sigma$ confidence intervals for some parameters are presented on Table \ref{table:results}. The standard value $N_{\mathrm{eff}} = 3$ is marginally ruled out with $99\%$ of confidence in all considered scenarios.

\section{Joint analysis with full CMB}

In spite of the above results, which show a better agreement between the Planck and local values of the Hubble-Lema\^itre parameter when an extra relativistic degree of freedom is added to the cosmic inventory, the performed tests do not involve the full CMB spectrum of anisotropies, but only a joint analysis of distance ladders as the CMB acoustic scale (that is, the position of the first peak in the anisotropy spectrum), the characteristic scale distances to the BAO peaks, and the luminosity distances to type Ia supernovae. The fit of the full CMB data with the $\Lambda$CDM model leads, in fact, to lower values of $N_{\text{eff}}$ and $H_0$ \cite{chineses,heavens}. In addition, higher values of $N_{\text{eff}}$ usually require higher values for the scalar spectral index $n_s$, which challenges the standard inflationary models \cite{heavens2,micol,micol2}. In this section we present a joint analysis of the full CMB data, obtained with the CosmoMC engine \cite{cosmomc,cosmomc2,camb} and the Planck 2015 likelihood \cite{Planck}, including the probes used above, namely the JLA compilation of SNe Ia \cite{sne:betoule14} and the Gaussian priors given by Verde {\it et al.} to the BAO scale $r_d$ \cite{Verde}, by Cooke {\it et al.} to the physical baryon density \cite{Cooke}, and by Riess {\it et al.} to the local value of $H_0$ \cite{Riess}. Our analysis was performed only for the $\Lambda$CDM model, while the generalised Chaplygin gas case will be presented in a forthcoming publication. The resulting probability density functions for some free and derived parameters are shown in Fig. 2, together with the $1\sigma$ and $2\sigma$ 2D confidence regions. Dashed lines refer to public available 2015 Planck chains (plikHM\_TT\_lowTEB and plikHM\_TT\_lowTEB\_post\_JLA), and solid lines to the joint analysis, plikHM\_TT\_lowTEB\_JLA including the aforementioned priors. The corresponding confidence intervals are shown in Table III. As expected, there is a positive correlation between $N_{\text{eff}}$ and $H_0$, as well as between $N_{\text{eff}}$ and $n_s$. The joint analysis marginally rules out the standard scenario of $3$ relativistic species, with $99\%$ of confidence. In contrast, an additional species is marginally within the $2\sigma$ confidence interval. The scalar spectral index results to be $n_s \approx 0.99$, close to a Harrison-Zeldovich spectrum but still in the region $n_s<1$ allowed by the inflation paradigm.

\section{Concluding Remarks}

The above results show that overcoming the $H_0$ tension between the CMB and HST observations may require a number of relativistic species that corroborates current experimental results in the neutrinos section of the standard model of particle physics. Indeed, the obtained best-fit $N_{\mathrm{eff}} \approx 4$ might be a clear signature of an additional, sterile, neutrino's family. We should stress, however, that the analysis we have performed includes only background tests, involving measurements of angular diameter and luminosity distances. The number of relativistic species also affects observations in the perturbative sector of cosmology, because the ratio between the matter and radiation densities defines, for example, the turnover of the matter power spectrum through the horizon scale value at the time of matter-radiation equality. Although the data do not determine this turnover precisely enough, a joint analysis of background and LSS observations would be complementary to the present results. Furthermore, a sterile neutrino with $1$eV mass would contribute with $\approx 8\%$ of a warm component in the present dark matter \cite{Dodelson}. On the other hand, despite the possibility presented here of conciliating the CMB acoustic scale with local $H_0$ measurements by adding a relativistic degree of freedom, the best-fit of the full CMB spectrum with the $\Lambda$CDM model in fact leads to a lower value of $N_{\mathrm{eff}}$ \cite{chineses,heavens}. It is also worth of note that a higher $N_{\mathrm{eff}}$ may be correlated to a higher spectral index of primordial fluctuations \cite{heavens2,micol,micol2}. The tests performed here, using distance rulers that are approximately model-independent, are complementary to other constraints, but the definite value of $N_{\mathrm{eff}}$ remains a subject for further investigation.

\begin{figure*}
 \begin{center}
 \includegraphics[width=0.9\textwidth]{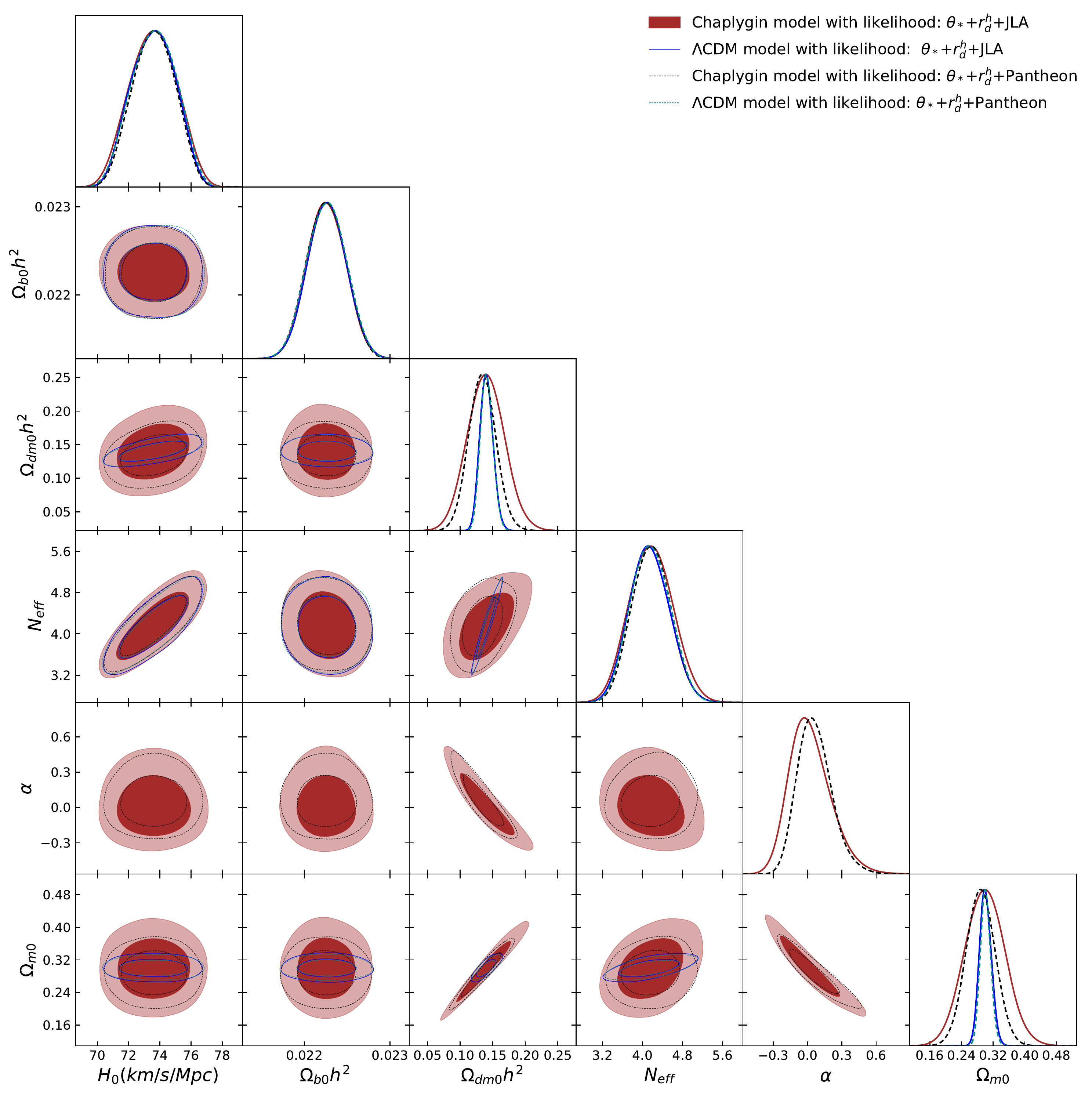} 
  \end{center}
  \caption{Probability distribution functions and marginalised confidence regions for our free parameters, for both the $\Lambda$CDM model and generalised Chaplygin gas.}
 \label{bxm}
 \end{figure*}

\begin{table*}
	\centering
\begin{tabular}{l|cc|cc|cc|cc|cc}
\hline
Model \& Data & $\chi^2_{\mathrm{bf}}$ & $\chi^2_{\nu}$ & $H_0^{\mathrm{bf}}$&$\left<H_0\right>\pm2\sigma$&$N_{\mathrm{eff}}^{\mathrm{bf}}$&$\left<N_{\mathrm{eff}}\right>\pm2\sigma$&$\Omega_{m0}^{\mathrm{bf}}$&$\left<\Omega_{m0}\right>\pm2\sigma$&$\alpha^{\mathrm{bf}}$&$\left<\alpha\right>\pm2\sigma$\\ [1ex]
\hline
{\bf Chaplygin} & & & &  & & & & & \\
 \multirow{2}{.9in}{$\theta_*$+$r_d^h$+JLA}  & {\blue 683.04} & {\blue 0.927}   &        73.37 & $73.59^{+2.81}_{-2.83}$ &             4.02 & $4.17^{+0.85}_{-0.80}$  &                  0.30 & $0.30\pm0.10$           & -0.04 & $0.02^{+0.41}_{-0.31}$ \\[1ex]
       & {\blue 683.22} & {\blue 0.926 } & {\blue 68.89}     & {\blue $70.22^{+1.63}_{-1.37}$} &            \multicolumn{2}{c|}{\blue $3.046$ (fixed)}  &  {\blue 0.30 }                 &     {\blue $0.26\pm0.08$ }       &{\blue  -0.01 } & {\blue $0.09^{+0.36}_{-0.28}$ } \\[1ex]
\cline{2-11}       
 \multirow{2}{.9in}{$\theta_*$+$r_d^h$+Pantheon} & {\blue 1026.89} &  {\blue 0.983} &    72.63 & $73.62^{+2.51}_{-2.55}$ &             3.91 & $4.17^{+0.74}_{-0.72}$ &                  0.30 & $0.29\pm0.07$         &  0.00 & $0.06^{+0.33}_{-0.26}$ \\[1ex]
       & {\blue 1026.96}  &  {\blue 0.982 } & {\blue 68.74 }     & {\blue $70.1^{+1.38}_{-1.09}$ } &            \multicolumn{2}{c|}{\blue $3.046$ (fixed)}  &  {\blue 0.29 }                 &     {\blue$0.27\pm0.06$ }       &{\blue 0.02 } & {\blue $0.04^{+0.27}_{-0.22}$ }  \\[1ex]
 \hline
 {\bf $\Lambda$CDM} & & &  & & & & & \\
 \multirow{2}{1in}{$\theta_*$+$r_d^h$+JLA}  & {\blue 682.93 }   &    {\blue 0.925} &     74.74 & $73.59^{+2.57}_{-2.62}$ &             4.29 & $4.14^{+0.78}_{-0.74}$ &                  0.30 & $0.30\pm0.03$          &  \multicolumn{2}{c}{\multirow{2}{*}{\blue 0 (fixed)} }                       \\[1ex]
       & {\blue 683.19}  &  {\blue 0.924 } &{\blue 69.13 }     & {\blue $70.33^{+1.53}_{-1.17}$ } &            \multicolumn{2}{c|}{\blue $3.046$ (fixed)}  &  {\blue 0.29}                 &     {\blue $0.28\pm0.02$ }     & \multicolumn{2}{c}{}  \\[1ex]
\cline{2-11}           
 \multirow{2}{1in}{$\theta_*$+$r_d^h$+Pantheon} & {\blue 1026.88}   &  {\blue 0.982 } &   74.11 & $73.64^{+2.61}_{-2.68}$ &             4.27 & $4.16^{+0.75}_{-0.72}$ &                  0.30 & $0.30^{+0.03}_{-0.02}$ &   \multicolumn{2}{c}{\multirow{2}{*}{\blue 0 (fixed)} }         \\[1ex]
        & {\blue 1026.89}   & {\blue 0.981 } & {\blue 68.97}     & {\blue $70.07^{+1.46}_{-1.12}$ } &            \multicolumn{2}{c|}{\blue $3.046$ (fixed)}  &  {\blue 0.30}                 &     {\blue $0.28\pm0.02$ }     &  \multicolumn{2}{c}{}   \\[1ex]
\hline
\end{tabular}
\caption{Best-fit values and 2$\sigma$  regions of some cosmological parameters for generalised Chaplygin gas and $\Lambda$CDM models using both JLA and Pantheon SNe compilation sets.}\label{table:results}
\end{table*}

\begin{figure*}
 \begin{center}
 \includegraphics[width=0.9\textwidth]{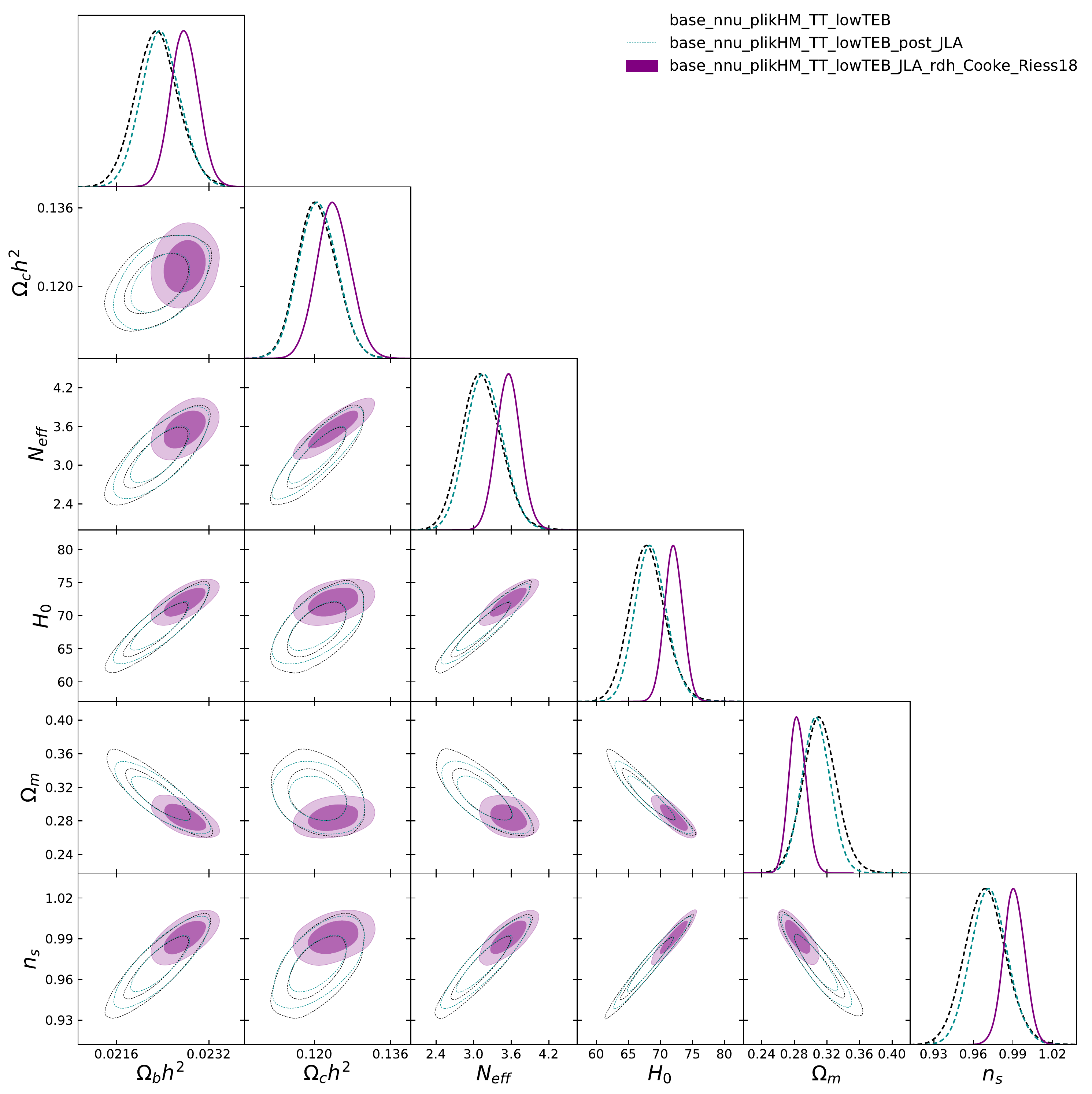} 
  \end{center}
  \caption{Probability distribution functions and marginalised confidence regions for some free and derived parameters of the $\Lambda$CDM model. The dashed lines were obtained with the public available 2015 Planck chains, while the solid curves and filled regions were obtained running CosmoMC with the plikHM\_TT\_lowTEB and JLA likelihoods, adopting the Gaussian priors for $r_d^h$, $\Omega_bh^2$ and $h$ previously mentioned in this work.}
 \label{bxm2}
 \end{figure*}

\begin{table*}
	\centering
    \begin{tabular}{lcccc}
    \hline
 parameter               & mean     & $1\sigma$             & $2\sigma$             & $3\sigma$               \\[1ex]
\hline
 $\Omega_bh^2$           & $0.0228$ & $\pm0.0002$           & $\pm0.0005$           & $ ^{+0.0008}_{-0.0007}$ \\[1ex]
 $\Omega_ch^2$           & $0.124$  & $ ^{+0.004}_{-0.003}$ & $\pm0.007$            & $ ^{+0.011}_{-0.010}$    \\[1ex]
 $100\theta_{MC}$        & $1.0406$ & $\pm0.0005$           & $\pm0.0010$            & $ ^{+0.0017}_{-0.0016}$ \\[1ex]
 $\tau$                  & $0.10$    & $ \pm0.02$  & $ \pm 0.04$ & $ \pm 0.06$   \\[1ex]
 $N_{\mathrm{eff}}$               & $3.556$  & $ ^{+0.189}_{-0.192}$ & $ ^{+0.379}_{-0.372}$ & $ ^{+0.610}_{-0.551}$    \\[1ex]
 ${\rm{ln}}(10^{10} A_s)$ & $3.15$  & $ \pm 0.04 $  & $\pm0.08$            & $ ^{+0.11}_{-0.12}$    \\[1ex]
 $n_s$                   & $0.991$  & $\pm0.008$            & $\pm0.016$            & $\pm0.025$              \\[1ex]
 \hline
 $H_0$                   & $72.05$ & $ ^{+1.45}_{-1.39}$ & $ ^{+2.77}_{-2.78}$  & $ ^{+4.40}_{-4.20}$   \\[1ex]
 $\Omega_m$              & $0.284$  & $ ^{+0.011}_{-0.010}$  & $ ^{+0.021}_{-0.019}$ & $ ^{+0.032}_{-0.028}$   \\[1ex]
 $\sigma_8$              & $0.858$  & $\pm0.019$            & $ ^{+0.038}_{-0.036}$ & $ ^{+0.056}_{-0.058}$   \\[1ex]
\hline
\end{tabular}

\caption{Mean values and credible intervals of the free cosmological parameters and some derived ones for the $\Lambda$CDM model, using the full CMB likelihood with the JLA SNe compilation set and $r_d^h$, $\Omega_bh^2$ and $h$ priors.}\label{table:results2}
\end{table*}

\section*{Acknowledgements}

We are thankful to J.S. Alcaniz and G. Gambini for helpful suggestions. Work partially supported by CNPq (SC, grant no. 307467/2017-1 and PCH, grant no. 310952/2018-2) and FAPESP (grant no. 2014/19164-6).

\end{document}